\documentclass[a4paper]{panl}
\usepackage{cite}
\usepackage{wrapfig}
\usepackage{graphicx}
\usepackage{amssymb}
\usepackage{amsfonts}
\usepackage{amsmath}
\usepackage{longtable}
\usepackage{rotating}
\usepackage{lscape}
\usepackage{epsfig}
\usepackage{multirow}

\usepackage{color}   
\usepackage{hyperref}
\hypersetup{
    colorlinks=true, 
    linktoc=all,     
    linkcolor=blue,  
    urlcolor=blue,
    citecolor=blue,
    linktocpage,
}

\originalTeX
\begin{document}

\title{Horizon Thermodynamics on Nice Slices of the Causal Diamond}
\maketitle
\authors{I.\,Nagle$^{a,}$\footnote{E-mail: ianagle@gmail.com}}
\setcounter{footnote}{0}
\from{$^{a}$\,Macquarie University}
\from{$^{a}$\,Sydney, Australia}

\begin{abstract}
\vspace{0.2cm}

\end{abstract}
\vspace*{6pt}
There is a deep link between gravity and thermodynamics; in a precise way gravity can be derived from entanglement entropy in conformal field theories. However, this depends crucially on properties of horizons, and asymptotic symmetries of phase space. To explore how this relation behaves under dynamical processes, we consider covariant gravitational phase space enhanced with bulk conformal symmetry. As is well known, the Noether-Wald entropy has an explicit form in terms of the Abbott-Deser-Tekin conserved surface charges of gauge theories. We find a new vector contribution to the Abbott-Deser-Tekin charges that arises for conformal symmetries. In applying this to the causal diamond, we derive a general relation for surface gravity, based on the conformal invariance of horizons, that allows us to find slices where the zeroth law holds, as well as the degree to which a first law arises on the phase space. 
\noindent

\tableofcontents

\label{sec:introduction}
\section{Introduction and Motivation}

The relation between gravity, entropy, and quantum field theories is one of the ongoing mysteries of modern physics. An important piece of this puzzle was the discovery, within the context of the AdS/CFT correspondence \cite{maldacena1}, that it was possible to derive the linearized field equations of gravity from the entanglement entropy of a conformal field theory \cite{metal1}. This observation has been abstracted into the proposal of ``entanglement equilibrium'' \cite{jacobson2} \cite{jacobsonvisser1}, which posits that the Einstein equations arise in causal diamonds in maximally symmetric spacetimes (de Sitter, anti-de Sitter, and Minkowski) by balancing contributions to Bekenstein's generalized entropy \cite{bekenstein1}. The arena where this acts is the classical phase space, where one can covariantly construct conserved quantities such as the entropy so long as the spacetimes in question are imbued with the requisite symmetries. 

For the causal diamond geometry in a maximally symmetric spacetime, a key step is that one can interpret the phase space in the gravity sector in the form of a first law of thermodynamics. However, one of the current limitations to this approach is that the identification of certain quantities such as the temperature, and the general form of contributions to the symplectic form, has primarily focused thus far on surface quantities associated with asymptotic Killing symmetry. This is logical as conserved charges in gauge theories are surface forms \cite{abbottdeser1} \cite{barnichbrandt1}, but in the case of the causal diamond \cite{jacobsonvisser1}, for instance, there are important bulk contributions to the first law that are proportional to the volume. 
\begin{equation}
T(\delta S - \eta \delta V) + \delta H_{matter} =0 
\end{equation}

Our particular route in this letter is to begin investigating how the conformal Killing symmetry of the causal diamond affects our description of the standard black hole Killing vector setup of surface gravity and acceleration \cite{jk1} \cite{wald}. We first find a general expression for the surface gravity of a conformal Killing horizon, clarify the relation between surface gravity and acceleration, and use this to investigate the zeroth law. For the particular case of the causal diamond the CKV foliates the entire spacetime, and so a zeroth law holds on each foliation. We then extend the Abbott-Deser-Tekin surface charge construction \cite{abbottdeser1} \cite{desertekin1} \cite{tekin1} to accomodate conformal symmetries in order to find the general form of bulk volume contributions to the phase space. This allows us to investigate extension of the first law away from the moment-of-time symmetry slice of the causal diamond.

\label{sec:setup}
\section{Setup}
For simplicity we focus just on gravity. We use the action and line element 
\begin{align}
\delta L &= \int \sqrt{g}(R-2\Lambda) \\
ds^2 &= - \left(1-\frac{r^2}{L^2} \right) du dv + r^2 d\Omega^2 . 
\end{align}

We will consider a causal diamond on this spacetime. We define this as the region ${|u| \leq R, |v| \leq R}$. The bifurication 2-surface, where the past and future null horizons meet, is located at $u=-R, v=R$, while the moment-of-time symmetry surface is along $t=(u+v)/2=0$.

Our approach will not be to add counterterms to regulate $\delta L$. Rather we keep the boundary term in the variation
\begin{equation}
\delta L = E \delta \phi + d\Theta
\end{equation}
and carefully choose symmetries and boundary conditions so that when $E=0$ (on-shell), the action will be stationary.

\label{sec:symm}
\section{Symmetries and Conformal Killing Vectors}
Recall that a Killing symmetry of a spacetime metric satisfies 
\begin{align}
\mathcal{L}_\xi g_{ab} = \nabla_a \xi_b + \nabla_b \xi_a = 0. 
\end{align}
For such a symmetry the action is stationary to first order
\begin{align}
x^a &\rightarrow x^a + \epsilon \xi^a \\
L &\rightarrow L + \epsilon \int d\nu \dot{x}^a \dot{x}^b \mathcal{L}_\xi g_{ab} + \mathcal{O}(\epsilon^2). 
\end{align}

We will be interested in a slightly relaxed version of the above; conformal Killing symmetries. These instead satisfy
\begin{align}
\label{ckv}
\mathcal{L}_\chi g_{ab} = 2\alpha g_{ab}
\end{align}
where the conformal factor $2\alpha = \frac{2}{n}\nabla_c \chi^c$. The causal diamond we consider has the conformal Killing field 
\begin{align}
\chi = \frac{L}{\sinh(R/L)} \left[ \left(\cosh\left(\frac{R}{L}\right)-\cosh\left(\frac{r}{L}\right) \right)\partial_u + \left( \cosh\left(\frac{R}{L}\right) - \cosh\left(\frac{v}{L}\right) \right)\partial_v \right].
\end{align}

On the bifurication 2-surface of the conformal diamond, these relations reduce to Killing symmetry.  

Since the boundaries of the causal diamond are null, surface charges on the $t=0$ slice depend on exact, rather than conformal, Killing structure. So the action and phase space are preserved. However, the action is deformed in the interior. Off the moment-of-time symmetry slice the null horizon of the conformal Killing field remains null, of course, but the conformal factor is nonvanishing and no longer reduces to exact Killing symmetry in the interior.

\label{subsec:surfgrav}
\subsection{Surface Gravity and Dynamics} - 
A helpful observation is that the surface gravity defined on a null horizon is conformally invariant \cite{jk1}. One can define this analogously to a black hole horizon \cite{wald}:      
\begin{align}
\nabla_b (\chi^2) \equiv -2 \kappa \chi_b. 
\end{align}

The standard way to find a formula for this is to use the Frobenius condition. This implies one can foliate the spacetime manifold with a vector field $\chi$. For a Kerr black hole this is true only on the horizon, while for the causal diamond (and also planar de Sitter), the CKV foliates the entire spacetime.
\begin{align}
\chi_{[a}\nabla_b \chi_{c]} =0. 
\end{align}
With conformal Killing symmetry we find the general formula for surface gravity 
\begin{align}
(\kappa + \alpha)^2 = \frac{-1}{2} \left[(\nabla^a \chi^b)(\nabla_a \chi_b) -\alpha^2 n \right].
\end{align}
(Note: the conformal factor of a CKV is distinct from the conformal factor of its background spacetime.)\\

We can relate this expression to acceleration by contracting the Frobenius condition with itself:
\begin{align}
\chi_{[a}\nabla_b \chi_{c]}\chi^{[a}\nabla^b \chi^{c]}=0  
\end{align}
This tells us the physical interpretion is
\begin{align}
\label{redaccel}
\kappa + \alpha = a V. 
\end{align}
These relations just depend on the metric and symmetries – so they are geometrical and also apply to, for instance, higher curvature gravity. Here we use the definitions $n=$dim, $u^2=-1$, $a^b=u^a \nabla_a u^b$, $a^2=a^b a_b$, and redshift $V^2 = -\chi^2$.

\label{subsec:zeroth}
\subsection{Zeroth Law for the Causal Diamond} - 
In black hole thermodynamics \cite{jacobsonvisser1} \cite{wald} \cite{bch} one looks for a zeroth law of thermodynamics by finding surfaces of constant surface gravity. The surface gravity for a black hole with a Killing vector is equal to the redshifted acceleration, which is in turn linked to the Unruh-Hawking temperature \cite{unruh1}. In our case the redshifted acceleration is equal to a combination of the standard surface gravity and conformal factor, as in equation \ref{redaccel}. 

For a stationary black hole the zeroth law holds on the null horizon. 
For a causal diamond, we find that the redshifted acceleration is constant along time translations, while the non-redshifted acceleration ($a=(\kappa+\alpha)/V $) is constant along orbits of $\chi$. 
\begin{align}
\mathcal{L}_t(aV) &= t^c \nabla_c (aV)=0 \\
\mathcal{L}_\chi a &= \chi^c \nabla_c a = 0. 
\end{align}
The non-redshifted acceleration preserves the symmetry of the diamond, while the redshifted acceleration is independent of its structure. 

\label{sec:charges}
\section{Covariant Phase Space}
Here we briefly review the covariant phase space formalism \cite{compere1} \cite{harlowwu1} \cite{iw1}, which is used to construct conserved quantities from symmetries on the phase space. 

The idea behind the covariant phase space formalism is to construct the symplectic form from the boundary term that appears in varying the action
\begin{align}
\delta L = E \delta \phi + d \Theta(\phi,\delta \phi). 
\end{align}
The usual variation $\delta$ is viewed as an exterior derivative on the field configuration space, which we take to commute with the exterior derivative $[d,\delta]=0$. From this one builds the pre-symplectic form
\begin{align}
\omega(\phi, \delta_1 \phi, \delta_2 \phi) \equiv \delta \Theta = \delta_2 \Theta_1 - \delta_1 \Theta_2. 
\end{align}
Hamiltonian charges arise by integrating the symplectic form over a Cauchy slice
\begin{align}
\delta H = \Omega = \int_\Sigma \omega.
\end{align}
In order for the phase space to be well-defined, one must choose deformations (which need not be small) and symmetries such that the action is stationary and the symplectic form closed ($d\Omega=0$). 
\begin{align}
\delta_1 g_{ab} &= h_{ab} \equiv g_{ab} - \bar{g}_{ab} \\
\delta_2 g_{ab} &= \mathcal{L}_\chi g_{ab} = 2 \alpha g_{ab}.
\end{align}



\label{subsec:surfbulk}
\subsection{Surface Charges and Bulk Flux} -

To gain insight on bulk dynamics, we find exact expressions for surface charges and bulk flux. As a useful technique, we will use an Abbott-Deser-Tekin \cite{abbottdeser1} \cite{desertekin1} \cite{tekin1} superpotential expansion of the linearized equations of motion to find general exact expressions for the surface and bulk quantities that arise with conformal symmetry, which we then connect back to the covariant approach through Noether's second theorem. 

The surface charge reproduces the standard Abbott-Deser expression, which both reduces to the ADM mass in flat space, and generalizes it to anti-de Sitter and de Sitter. The general expression for bulk flux has not appeared in the literature before, to my knowledge, and arises due to bulk conformal symmetry. The form of the surface charge and bulk flux are uniquely determined by the deformations $\delta_1 g_{ab}$ and $\delta_2 g_{ab}$ on phase space, which must satisfy the linearized equations of motion.

We first recall Noether's second theorem, which says that for a continuous diffeomorphism
\begin{align}
\frac{\delta L}{\delta \phi} \delta_\chi \phi &= d S_\chi .
\end{align}
Appling to the linearized equations of motion gives the form (here $\epsilon = n \wedge \tilde{\epsilon}$)
\begin{align}
\delta S_\chi &=  \left(\frac{-\tilde{\epsilon} }{8 \pi G} \delta E^{ab}\chi_b \right). 
\end{align}
Our plan is to write $\delta E^{ab}\chi_b$ in terms of a (to be defined) conserved surface charge $d k_\chi$ and bulk flux vector $l_\chi$. As an intermediate step we write the linearized equations of motion in terms of a superpotential.
\begin{align}
\delta E^{ab} = \nabla_c \nabla_d K^{acbd} + Y^{ab}.
\end{align}
The $Y^{ab}$ term is unimportant and vanishes on-shell. The superpotential is 
\begin{align}
K^{acbd} &= \frac{1}{2} \left[ g^{ad} \tilde{h}^{bc} + g^{bc} \tilde{h}^{ad} - g^{ab} \tilde{h}^{cd} - g^{cd} \tilde{h}^{ab} \right], ~ \text{with} ~ 
\tilde{h}^{ab} = h^{ab} - \frac{1}{2}\bar{g}^{ab}.
\end{align}

Contracting with a conformal Killing vector allows one to write 
\begin{align}
\chi_b \delta E^{ab} &= \nabla_c \left[ \chi_b \nabla_d K^{acbd} - K^{adbc} \nabla_d \chi_b \right] \\
&+ \left(g_{cb} \nabla_d \alpha + g_{bd} \nabla_c \alpha - g_{cd} \nabla_b \alpha \right) K^{acbd}.
\end{align}

To connect this to the symplectic form in covariant phase space, we can use Noether's second theorem, nilpotency of the action under $\delta^2$, and the Cartan formula to obtain
\begin{align}
\omega = -\chi \cdot \epsilon E^{ab} h_{ab} + \delta S_\chi + d B.
\end{align} 
Imposing the equations of motion and integrating over a Cauchy slice
\begin{align}
\Omega &= \delta H_\chi = \int_\Sigma \delta S_\chi \\
&= \int_{\partial \Sigma} k_\chi + \int_\Sigma l_\chi.
\end{align} 
In the last line we have split up the linearized term $\delta E^{ab} \chi_b$ into an antisymmetric component supported on the boundary, and a symmetric part that is only nonvanishing in the bulk. We denote these as the surface charge $k_\chi$ and vector flux $l_\chi$.

\label{subsec:first}
\subsection{First Law on Nice Slices} - 

We now apply the above to the causal diamond. The idea at this moment is to investigate to what extent the first law holds covariantly, off the $t=0$ slice. 

To isolate a temperature dependence we need to factor the acceleration (and thus Unruh temperature) from the surface charge and the bulk flux. We here focus on the bulk flux. 
\begin{align}
\int_\Sigma l_\chi = \int \epsilon_a \left(g_{cb} \nabla_d \alpha + g_{bd} \nabla_c \alpha - g_{cd} \nabla_b \alpha \right) K^{acbd}
\end{align}
On surfaces where either the acceleration $a$ or redshifted Unruh temperature $aV$ are constant, we have $\nabla_c \alpha = - \nabla_c \kappa$. Thus, one can factor the surface gravity (as opposed to the Unruh temperature), when $n^c \nabla_c \kappa \approx \kappa $, along some suitable direction $n^c$. 

For the causal diamond the conformal factor takes the schematic form $2 \alpha = \gamma \sinh \frac{u+v}{2L}$. Thus 
\begin{align}
n^c \nabla_c \alpha \approx \gamma \left( \sinh \frac{u+v}{2L} + e^{-(u+v)/2L}  \right).
\end{align}
So one can identify this approximately with the surface gravity. Collecting terms, this gives an approximate first law

\begin{align}
\delta H_\chi^{grav} &= \int_\Sigma k_\chi + \int_{\partial \Sigma} l_\chi \\
&= \kappa \delta A + [\kappa + \mathcal{O}(e^{-(u+v)/2L})] \eta \delta V . 
\end{align}

\label{sec:conclusions}
\section{Conclusions}

In this letter, we have clarified the relation between surface gravity and acceleration for spacetimes with conformal Killing symmetry, and found surface gravity formulas that extend deep into the bulk. By considering the Abbott-Deser-Tekin superpotential approach to deformations of the linearized equations of motion, we have obtained a general covariant expression for the volume-law behavior of bulk flux, which is fixed based on deformations and symmetries that preserve the symplectic form. This gives us the ability to precisely investigate zeroth and first laws off the bifurication 2-surface and $t=0$ slice, which we report preliminary investigation of. Since the first law off the moment-of-time symmetry surface holds only approximately for the causal diamond, it would be interesting to see if there are situations where this affects the `gravitation from CFTs' connection of \cite{metal1}. As other spacetimes, such as planar de Sitter, also admit CKVs, it may be possible to investigate the conformal horizon thermodynamics of such metrics. 

\label{sec:ack}
\section*{Acknowledgements}
It is a pleasure to thank Peter Bouwknegt, Brett Parker, Alexei Gilchrist, Shira Chapman, and Victor Kim for useful conversations.

\label{sec:funding}
\section*{Funding}
This work was supported by ongoing institutional funding. No additional grants to carry out or direct this particular research were obtained.

\label{sec:conflict}
\section*{Conflict of interest}
The author declares no conflicts of interest.

\label{sec:pub}
\section*{Publication notice}
This a preprint of the Work accepted for publication in Physics of Particles and Nuclei Letters, ©, copyright 2025, Pleiades Publishing, Ltd.; \href{https://link.springer.com/journal/11497/volumes-and-issues/22-1}{https://link.springer.com/journal/11497/volumes-and-issues/22-1} 


\end{document}